\documentclass[]{elsarticle}
\usepackage{amsmath, amsthm, amssymb}
\usepackage[margin=0.8in]{geometry}
\usepackage{subfig}

\title{Positivity-preserving scheme for two-dimensional advection-diffusion equations including mixed derivatives}

\author[yrk,ccfe]{E.J. du Toit\corref{cor1}}
\ead{ejdt500@york.ac.uk}
\author[ccfe]{M.R. O'Brien}
\ead{}
\author[yrk]{R.G.L. Vann}
\ead{}

\address[yrk]{York Plasma Institute, Department of Physics, University of York, York, YO10 5DD, UK}
\address[ccfe]{Culham Centre for Fusion Energy, Abingdon, OX14 3DB, UK}

\cortext[cor1]{Corresponding author}

\date{}                     

\begin{document}

\begin{abstract}
In this work, we propose a positivity-preserving scheme for solving two-dimensional advection-diffusion equations including mixed derivative terms, in order to improve the accuracy of lower-order methods. The solution to these equations, in the absence of mixed derivatives, has been studied in detail, while positivity-preserving solutions to mixed derivative terms have received much less attention. A two-dimensional diffusion equation, for which the analytical solution is known, is solved numerically to show the applicability of the scheme. It is further applied to the Fokker-Planck collision operator in two-dimensional cylindrical coordinates under the assumption of local thermal equilibrium. For a thermal equilibration problem, it is shown that the scheme conserves particle number and energy, while the preservation of positivity is ensured and the steady-state solution is the Maxwellian distribution. 
\end{abstract}

\maketitle

\section{Introduction}
Two dimensional advection-diffusion equations have widespread applications in physics, engineering and finance, and can generally be written as
	\begin{equation}
	\label{eq:pde}
	u_t = A u_{xx} + B u_{xy} + C u_{yy} + D u_x + E u_y + F u
	\end{equation}
where $u = u(x,y,t)$. As these equations are often too difficult to solve analytically, numerical solutions are required. For $F = 0$, these equations can be written in a two-dimensional advection-diffusion form,
	\begin{equation}
	\label{eq:pde:2}
	\frac{\partial u}{\partial t} = \nabla \cdot \left( - \vec{a} u + \hat{k} \cdot \nabla u \right)
	\end{equation}
where $u = u(x,y,t)$ is advected by the $2$D vector $\vec{a}(x,y,t)$ and diffused by the tensor $\hat{k}(x,y,t)$. If the initial condition $u(x,y,0) \ge 0$ for all $(x,y)$, then the solution must always be positive, i.e. $u(x,y,t) \ge 0$ for all $(x,y,t)$. A good numerical method will also preserve the monotonicity of the initial condition, but these conditions poses a particular challenge if a change of coordinates, in order to eliminate the mixed derivative terms $u_{xy}$ throughout $(x,y)$ space, is not possible. 

A particular application of two-dimensional advection-diffusion equations is the Fokker-Planck collision operator, which can typically be written in the form (\ref{eq:pde:2}) and has a wide range of applications in plasmas in the laboratory (e.g. magnetic and inertial thermonuclear fusion), space (e.g. Earth's magnetosphere), and astrophysics (e.g. solar coronal mass ejections) \cite{Taitano_2015}. Positivity-preserving solutions to two-dimensional advection and diffusion equations, in the absence of mixed derivative terms, have been studied in detail \cite{Fazio_2009, Hundsdorfer_1995}, but solutions where mixed derivative terms are present have received much less attention. Recent research have therefore focused on developing improved and refined higher-order methods for solving advection-diffusion equations, as higher-order methods are typically more accurate, but also more complex, than lower-order methods, which are more reliable and robust \cite{Huynh_2007}. 

In this paper, we propose a scheme for improving the accuracy of lower-order methods, in particular with respect to the preservation of positivity, when solving two-dimensional advection-diffusion equations in the presence of mixed derivatives. The proposed scheme is applicable to any advection-diffusion equation of the form (\ref{eq:pde:2}), as well as equations of the form (\ref{eq:pde}) where the solution must remain positive. It is applied to a two-dimensional diffusion equation with mixed derivatives, for which the analytical solution is known, as well as the Fokker-Planck collision operator in cylindrical coordinates.

In our proposed scheme, in order to preserve positivity, the mixed derivative term is rewritten as an advective equation, for which many positivity-preserving solutions exist \cite{Fazio_2009, Hundsdorfer_1995, Huynh_2007, Arber_2002, Fijalkow_1999, Laney_1998}. We show that, compared to central finite-difference methods, this scheme has the same order of accuracy, while ensuring the preservation of positivity.

One-dimensional solutions to the Fokker-Planck collision operator have been studied for decades, but detailed numerical discretizations in two-dimensions, particularly in cylindrical coordinates where mixed derivative terms are present, have only recently been studied. A numerical approximation to the Fokker-Planck collision operator should ensure the conservation of particle number, momentum and energy and the preservation of positivity, and ensure a steady-state Maxwellian distribution under thermal equilibration. The most successful approaches consist of an extension of the one-dimensional Chang and Cooper scheme \cite{Chang_1970} to two-dimensions, but this method does not guarantee the preservation of positivity if the solution is far from equilibrium \cite{Yoon_2014}. A fully implicit finite element algorithm, using appropriate flux limiters to ensure the preservation of positivity, the conservation of particle number, momentum, and energy, has also been developed, but is intensive \cite{Taitano_2015}. A good review of other numerical methods can also be found in \cite{Taitano_2015}.

Here, we present an alternative approach to the Fokker-Planck collision operator in two-dimensional cylindrical coordinates. The solution is based on the proposed positivity-preserving scheme and extends the Chang and Cooper scheme, based on the assumption of local thermal equilibrium, to ensure an accurate steady-state solution is obtained. We approximate the collision operators by assuming the distribution collides with a background Maxwellian distribution, and it is shown that, if this approximation holds, thermal equilibration occurs at the theoretically predicted rate. The proposed scheme conserves particle number and energy, while the preservation of positivity is ensured.

The paper is structured as follows: Section $2$ introduces our proposed positivity-preserving scheme for solving two-dimensional advection-diffusion equations, including the treatment of mixed derivative terms. Section $3$ discusses the Fokker-Planck collision operator in cylindrical coordinates, including the assumption of local thermal equilibrium and thermal equilibration tests, and is followed by a short summary in Section $4$.

\section{$2$D diffusion with mixed derivatives}
Positivity-preserving solutions to two-dimensional advection and diffusion equations have been studied in detail \cite{Fazio_2009, Hundsdorfer_1995}, but solving mixed derivative terms have received much less attention. The reason for this is that typically a change of coordinate system can be performed in order to eliminate the mixed derivative terms, or the mixed derivative terms are weak compared to the advection-diffusion terms and can therefore be neglected. If this is not possible, however, a positivity-preserving solution is required for the unmodified equation.

Consider, as an example, the two-dimensional diffusion equation
	\begin{equation}
	\label{eq:example}
	u_t = u_{xx} + u_{xy} + u_{yx} + u_{yy}
	\end{equation}
with initial condition
	\begin{equation*}
	u(x,y,t=0) = \exp{[-x^2 - y^2]}
	\end{equation*}
and open boundary conditions, such that the grid on which we solve $u(x,y,t)$ must be large enough to ensure $u = 0$ at the boundaries always. Typically, this equation will be solved by performing a change of coordinates in order to eliminate the mixed derivative terms $u_{xy}$ and $u_{yx}$. In this way, an analytical solution can be obtained,
	\begin{equation}
	U(x,y,t) = \frac{1}{\sqrt{1 + 8t}} \exp{\left[-\left(\frac{1}{2}x^2 + \frac{1}{2}y^2 - xy \right) - \frac{1}{1 + 8t} \left(\frac{1}{2}x^2 + \frac{1}{2}y^2 + xy \right) \right]}
	\end{equation}

The aim of this section is to introduce a positivity-preserving scheme for solving mixed derivative terms, thereby solving equation (\ref{eq:example}) without a change of coordinate system. The numerical solution $u_{i,j}$ is then compared to the analytical solution through an RMS error given by
	\begin{equation}
	E_\text{rms} = \sqrt{\frac{1}{N^2} \sum_{i,j} (u_{i,j} - U_{i,j} )^2}
	\end{equation}
where $N$ is the number of grid points in both the $x$- and $y$-directions and the labels $i,j$ refers to the $i^\text{th}$ node in the $x$-direction and the $j^\text{th}$ node in the $y$-direction.

Although higher-order methods for solving (\ref{eq:example}) can be used obtained with the use of flux limiters, these are often complicated and less robust than lower-order methods. Lower-order methods, on the other hand, tend to be less accurate, and, as will be shown, does not guarantee the preservation of positivity. Our proposed scheme improves the accuracy of lower-order methods, especially by ensuring the preservation of positivity. For this reason, we only consider two second-order accurate finite-difference methods. The diffusion terms $u_{xx}$ and $u_{yy}$ are straightforward to solve by taking second-order derivatives,
	\begin{equation}
	\label{eq:diffusion}
	u_{xx} = \frac{\partial}{\partial x}\frac{\partial u}{\partial x} = \frac{u_{i+1,j} - 2u_{i,j} + u_{i-1,j}}{\Delta x^2} + \mathcal{O}(\Delta x^2)
	\end{equation}
and is second-order accurate and preserves positivity.

The mixed derivatives can be solved with a similar central finite-difference method, where the boundary values are determined as an average, i.e.
	\begin{equation*}
	u_{i+1/2} = \frac{1}{2} \bigg( u_i + u_{i+1} \bigg)
	\end{equation*}
but, although such a method is second-order accurate, it does not guarantee positivity. 

Positivity-preserving schemes exist for linear advection equations, and we therefore aim to rewrite the mixed derivative terms as advection equations in order to employ these schemes and preserve positivity.

\subsection{Positivity-preserving approximation}
Positivity-preserving approximations for linear advection equations have been studied in detail \cite{Fazio_2009, Hundsdorfer_1995, Arber_2002, Fijalkow_1999, Laney_1998}, while such approximations for mixed derivative terms have received much less attention. As central finite-difference method solutions to mixed derivative equations do not preserve positivity, we rewrite these terms as advection equations, and then apply a positivity-preserving approximation. In order to rewrite the mixed derivative terms as advection equations, we define the function
	\begin{equation}
	\label{eq:log:derivative}
	v = \frac{1}{u} \frac{\partial u}{\partial y}
	\end{equation}
so that we can write
	\begin{equation}
	\label{eq:mixed:advection}
	u_{xy} = \frac{\partial}{\partial x} \frac{\partial u}{\partial y} = \frac{\partial}{\partial x} v \cdot u
	\end{equation}
where $v = v(x,y,u)$. This results in the mixed derivative term being written as a non-linear advection equation, which can be solved using second-order positivity-preserving schemes, as discussed in \ref{app:Hundsdorfer}. Higher-order methods exist, but as these methods are more complicated and less robust than lower-order methods, and as we aim to compare our numerical approximation to a second-order central finite-difference method, we use a second-order positivity-preserving scheme. The non-linearity can be resolved through Picard iteration, discussed later. A more straightforward solution is through Picard linearizing, where the function $v$ is determined from the known values of $u^{n}$ in order to find the solution $u^{n+1}$ at the next time step.

The function $1/u \to \infty$ as $u \to 0$, which can lead to singularities. In order to deal with such singularities, it is assumed that the grid is fine enough and the function smooth enough such that the difference between neighbouring nodes is small, such that $u_{j+1} - u_{j-1} \ll u_j$, and the function
	\begin{equation*}
	v = \frac{1}{u} \frac{\partial u}{\partial y} = \frac{1}{2\Delta y} \frac{u_{j+1} - u_{j-1}}{u_j} \approx 0
	\end{equation*}
if $u_j < \epsilon$, where $\epsilon \ll 1$. We therefore set $v = 0$ if $u_j < \epsilon$, where $\epsilon = 10^{-16}$. The value of $\epsilon$ will depend on the particular problem, as well as the relative values of $u$ and the coarseness of the grid. In the problems we consider, the solutions approach zero exponentially and $\epsilon = 10^{-16}$ is small enough.

\subsection{Time evolution}
The time evolution will be solved implicitly. For example, the diffusion equation (\ref{eq:diffusion}) is solved as
	\begin{equation}
	u_t = \frac{1}{\Delta t} \bigg[ u^{n+1} - u^n \bigg] = \frac{1}{\Delta x^2} \bigg[ u_{i+1,j}^{n+1} - 2u_{i,j}^{n+1} + u_{i-1,j}^{n+1} \bigg] + \mathcal{O}(\Delta t)
	\end{equation}
which is first-order accurate in time, unconditionally stable and independent of the Courant number \cite{Courant_1928}.

\subsection{Numerical results}
\begin{figure}[!hbt]
	\centering
		\subfloat[]{%
			\includegraphics[width=0.33\textwidth]{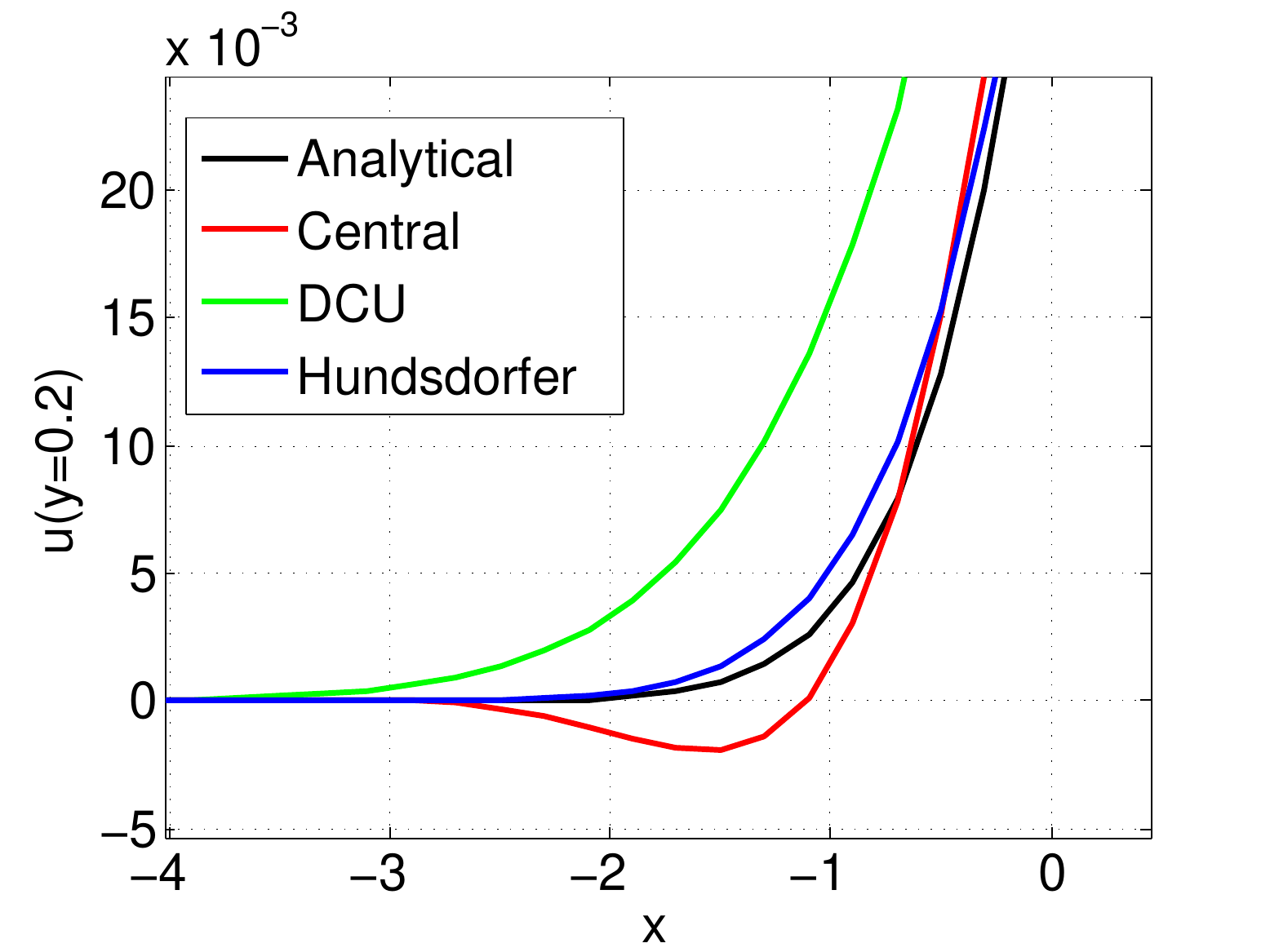}}
		\hfill
		\subfloat[]{%
			\includegraphics[width=0.33\textwidth]{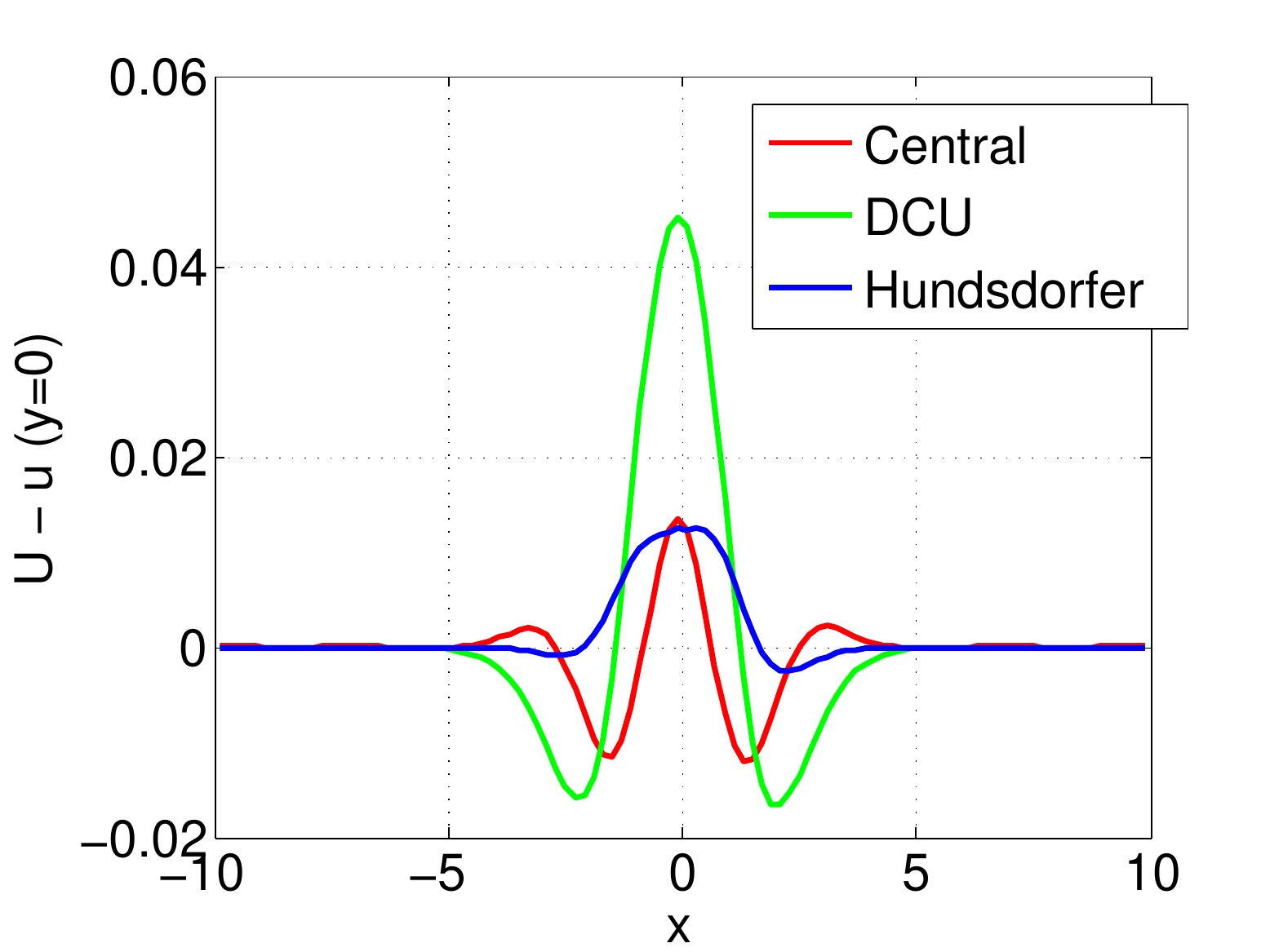}}
		\hfill
		\subfloat[]{%
			\includegraphics[width=0.33\textwidth]{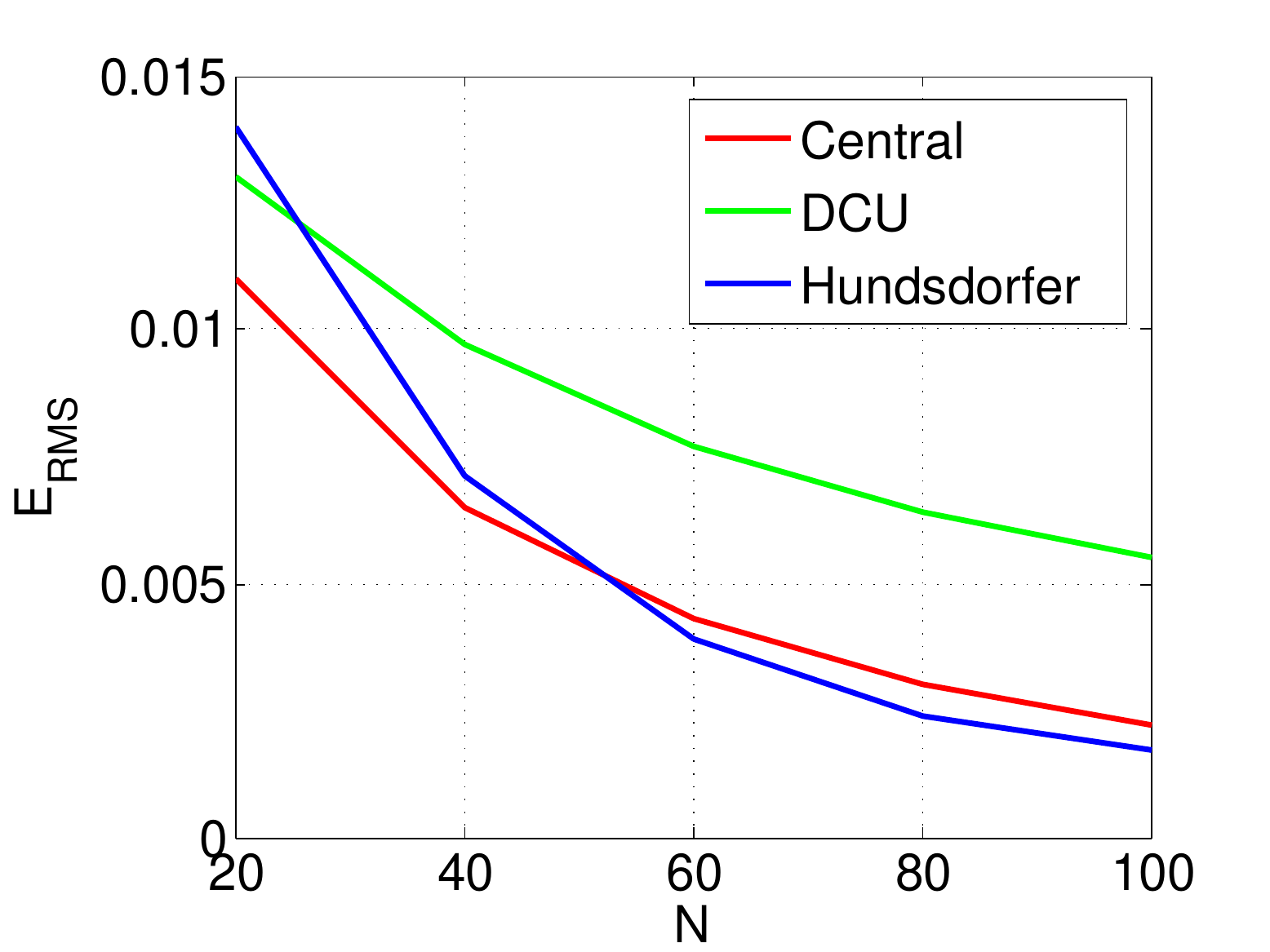}}
	\caption[]{(a) The comparison of analytical to numerical solutions shows that the central difference scheme does not preserve positivity. (b) The difference between the analytical and numerical approximations for $y = 0$, with $N = 100$ grid points. (c) The comparison of $E_\text{RMS}$ for different numerical methods, as a function of number of grid points $N$.}
	\label{fig:example:y}
	\end{figure}

We solve equation (\ref{eq:example}) on a uniform grid with size $x,y \in (-10,10)$ and $\Delta x = \Delta y = 20/N$, with $\Delta t = 0.1$ for $20$ timesteps. We use only a single Picard iteration, in which the coefficients $v$ are calculated from the known $u^n$ to obtain $u^{n+1}$. Comparisons between the analytical and numerical solutions are shown in figure \ref{fig:example:y}.

As expected, the central finite-difference scheme does not preserve positivity, while both the first-order donor-cell upwind (DCU) scheme and second-order Hundsdorfer scheme (\ref{app:Hundsdorfer}) preserve positivity. The accuracy of the central finite-difference and Hundsdorfer schemes are both second-order, while the DCU scheme is first-order accurate, as expected. For small $N$, the grid is very coarse, such that variations between neighbouring cells are great, and the Hundsdorfer scheme is effectively first-order as the flux-limiter $\phi \approx 0$ due to large variations between neighbouring nodes. When increasing $N$ the Hundsdorfer scheme improves to become second-order accurate.


\section{Fokker-Planck collision operator in cylindrical coordinates}
The Fokker-Planck collision operator describes the local collisional relaxation process of distribution functions in plasmas under the assumption of binary, small-angle collisions \cite{Rosenbluth_1957, Landau_1937, Chandrasekhar_1943} and is regarded, along with Vlasov and Maxwell's equations, as the basis for weakly-coupled plasmas in all collisionality regimes. It conserves particle number, momentum and energy, preserves positivity of the distribution function, and satisfies the Boltzmann H-theorem, such that the steady-state solution is given by the Maxwellian distribution function. Despite this, however, it is a stiff advection-diffusion operator in velocity space, and nonlinear when solving the collision operators using the Landau integral \cite{Landau_1937} or the Rosenbluth potentials \cite{Rosenbluth_1957}, which leads to several difficulties in dealing with it numerically \cite{Taitano_2015}.

In the presence of a strong magnetic field, the gyrofrequency exceeds all frequencies of interest, such that the dependence of the distribution function on the gyroangle can be neglected. The Fokker-Planck collision operator is normally solved in spherical coordinates $(v,\theta)$ where it contains no mixed derivative terms. In some cases, however, the distribution function must be solved under multiple effects, some of which may be best described in cylindrical coordinates $(v_\parallel,v_\perp)$, such as radiofrequency current drive \cite{OBrien_1986, Karney_1986, Maekawa_2012}, and an approximation to the collision operator in cylindrical coordinates is therefore required. The Fokker-Planck collision operator in cylindrical coordinates is therefore considered,
	\begin{equation}
	\label{eq:collisions:cylindrical}
		\begin{aligned}
	\frac{\partial f}{\partial t} = \frac{1}{v_\perp} \frac{\partial}{\partial v_\perp} v_\perp &\bigg[ D_{\perp \perp} \frac{\partial f}{\partial v_\perp} + D_{\perp \parallel} \frac{\partial f}{\partial v_\parallel} - F_\perp f \bigg] \\
	+ \frac{\partial}{\partial v_\parallel} &\bigg[ D_{\parallel \parallel} \frac{\partial f}{\partial v_\parallel} + D_{\parallel \perp} \frac{\partial f}{\partial v_\perp} - F_\parallel f \bigg]
		\end{aligned}
	\end{equation}
where $v_\parallel$ is the velocity parallel and $v_\perp$ is the velocity perpendicular to the magnetic field and $f = f(v_\parallel,v_\perp,t)$ is the electron distribution function. For simplicity, the collision operators $D$ and $F$ are taken to be those for an isotropic, background Maxwellian \cite{Karney_1986}. This assumption implies that energy will only be conserved if the distribution collides with a background Maxwellian; otherwise only particle number will be conserved, while the preservation of positivity is ensured by the numerical scheme. In general, the collision operators could be determined from the Landau integral \cite{Landau_1937} or the Rosenbluth potentials \cite{Rosenbluth_1957}, but this is not considered in this work.

The solution to the Fokker-Planck equation (\ref{eq:collisions:cylindrical}) is obtained using the positivity-preserving method discussed earlier, but using the first-order DCU scheme (\ref{app:Hundsdorfer}) for solving the advective equations. This simplifies the extension of the Chang and Cooper scheme, which ensures the correct equilibrium solution under the assumption of local thermal equilibrium. The result is a first-order solution that always conserves particle number and preserves positivity, while energy is conserved if the assumptions in calculating the collision operator are satisfied.

\subsection{$\delta$-splitting}
The equilibrium solution to the collision operator is a Maxwellian distribution function, which is a strongly (exponentially) varying function of $v$. In order to ensure the conservation of particle number, Chang and Cooper \cite{Chang_1970} introduced a weighted averaging scheme based on the assumption of local thermal equilibrium. We employ a similar scheme, but, as we employ a flux-conserving scheme for discretising the advection-diffusion equation, particle number is conserved by construction. In this case, the assumption of local thermal equilibrium ensures the correct steady-state distribution and greatly improves the accuracy of the time evolution of the distribution function.


In general, the time evolution of the distribution function can be solved under multiple effects, such as collisions, electric fields and plasma-wave interactions \cite{OBrien_1986, Karney_1986, Maekawa_2012}, and this method could easily be extended to include these additional terms. We will only consider the effect of collisions, for which the Fokker-Planck equation can be written as the divergence of a flux,
	\begin{equation*}
	\frac{\partial f}{\partial t} = \nabla \cdot \vec{S}_c
	\end{equation*}

If we assume steady-state, or that the distribution is in local thermal equilibrium, particle number is exactly conserved if the flux $\vec{S}_c = 0$. If we consider the $v_\perp$ part of the collision operator, i.e. the first term on the right-hand side of equation (\ref{eq:collisions:cylindrical}),
	\begin{equation*}
	\frac{\partial f}{\partial t} = \frac{1}{v_\perp} \frac{\partial}{\partial v_\perp} v_\perp \bigg[ D_{\perp \perp} \frac{\partial f}{\partial v_\perp} + D_{\perp \parallel} \frac{\partial f}{\partial v_\parallel} - F_\perp f \bigg]
	\end{equation*}
for which the flux is
	\begin{equation}
	\label{eq:flux}
	D_{\perp \perp} \frac{\partial f}{\partial v_\perp} + D_{\perp \parallel} \frac{\partial f}{\partial v_\parallel} - F_\perp \, f
	\end{equation}.

This can be rewritten, under the assumption of local thermal equilibrium, to obtain
	\begin{equation}
	\label{eq:flux2}
	\frac{\partial f}{\partial v_\perp} = \frac{1}{D_{\perp \perp}} \bigg( F_\perp - D_{\perp \parallel} \cdot g \bigg) f
	\end{equation}
where
	\begin{equation*}
	g = \frac{1}{f} \frac{\partial f}{\partial v_\parallel}
	\end{equation*}
is equivalent to equation (\ref{eq:log:derivative}). Equation (\ref{eq:flux2}) can be solved to obtain the solution
	\begin{equation}
	\label{eq:f_i+1}
		\begin{aligned}
	f_{i,j+1} &\sim f_{i,j} \exp \bigg[ \frac{ F_\perp - D_{\perp \parallel} \cdot g }{D_{\perp \perp}} \Delta v_\perp \bigg] \\
	&= f_{i,j} \exp \bigg[ A_g - B_g \bigg]
		\end{aligned}
	\end{equation}
where the labels $(i,j)$ refers to the $i^\text{th}$ node in the parallel direction and the $j^\text{th}$ node in the perpendicular direction. The functions $A_g$ and $B_g$ are defined as
	\begin{equation}
	\label{eq:delta:perp:AB}
	A_g = \frac{F_\perp}{D_{\perp \perp}} \Delta v_\perp \quad \quad \quad ; \quad \quad \quad B_g = \frac{ D_{\perp \parallel} \cdot g }{D_{\perp \perp}} \Delta v_\perp
	\end{equation}

For simplicity, the advective and mixed derivative terms are approximated using the first-order DCU scheme (see \ref{app:Hundsdorfer}). We then rewrite the $F_\perp$-term (for $F_\perp < 0$), as
	\begin{equation*}
		\begin{aligned}
	\frac{\partial f}{\partial t} = \frac{1}{\Delta v} \bigg[ &\left( \frac{v_\perp - \Delta v/2}{v_\perp} \right) F_\perp \left(v_\parallel, v_\perp - \frac{\Delta v}{2} \right) \delta_\perp f_{i,j} \\
	&- \left( \frac{v_\perp + \Delta v/2}{v_\perp} \right) F_\perp \left(v_\parallel, v_\perp + \frac{\Delta v}{2} \right) \delta_\perp f_{i,j+1} \bigg]
		\end{aligned}
	\end{equation*}
and solve for $\delta_\perp$ from equation (\ref{eq:flux}). We do this by considering the flux across the $(v_\perp + \Delta v/2)$-boundary, for $D_{\perp \parallel} \cdot g > 0$,
	\begin{equation*}
	\frac{1}{\Delta p} D_{\perp \perp,j+1/2} (f_{i,j+1} - f_{i,j}) + D_{\perp \parallel, j+1/2} \cdot g_{j+1/2} \cdot f_{i,j+1} - F_{\perp,j+1/2} \cdot f_{i,j+1} \cdot \delta_\perp = 0
	\end{equation*}
which can be used to obtain, along with (\ref{eq:f_i+1}),
	\begin{equation*}
	f_{ij} e^{A_g} e^{-B_g} - f_{ij} + B_g \, f_{ij} e^{A_g} e^{-B_g} - A_g e^{A_g} e^{-B_g} f_{ij} \, \delta_\perp = 0
	\end{equation*}
and therefore,
	\begin{equation}
	\delta_\perp = \frac{1}{A_g} \bigg( 1 + B_g - e^{-A_g} e^{B_g} \bigg)
	\end{equation}
with the same result obtained for the flux across the $(v_\perp - \Delta v/2)$-boundary.

For $D_{\perp \parallel} \cdot g < 0$ (and $F_\perp < 0$), the solution to $\delta_\perp$ is given by
	\begin{equation}
	\delta_\perp = \frac{1}{A_g} \bigg( 1 + B_g e^{-A_g} e^{B_g} - e^{-A_g} e^{B_g} \bigg)
	\end{equation}
with $A_g$ and $B_g$ given by (\ref{eq:delta:perp:AB}). A similar approach is used for the flux in the $v_\parallel$-direction in order to obtain solutions for $\delta_\parallel$, but in this case there are four possibilities,
	\begin{center}
	\begin{tabular}{ c | c || c }
	$F_\parallel$ & $D_{\parallel \perp} \cdot h$ & $\delta_\parallel$ \\
	\hline
	\hline
	$ > 0$ & $ > 0$ & $\frac{1}{A_h} \bigg( e^{A_h-B_h} + B_h e^{A_h-B_h} - 1 \bigg)$ \\
	\hline
	$ < 0$ & $ > 0$ & $\frac{1}{A_h} \bigg( 1 + B_h - e^{B_h-A_h} \bigg)$ \\
	\hline
	$ > 0$ & $ < 0$ & $\frac{1}{A_h} \bigg( e^{A_h-B_h} - 1 + B_h \bigg)$ \\
	\hline
	$ < 0$ & $ < 0$ & $\frac{1}{A_h} \bigg( 1 - e^{B_h-A_h} + B_h e^{B_h-A_h} \bigg)$ \\
	\end{tabular}
	\end{center}
with the functions $A_h$ and $B_h$ defined as
	\begin{equation*}
	A_h = \frac{F_\parallel}{D_{\parallel \parallel}} \Delta v \quad \quad \quad ; \quad \quad \quad B_h = \frac{ D_{\parallel \perp } \cdot h }{D_{\parallel \parallel}} \Delta v
	\end{equation*}
and
	\begin{equation*}
	h(p_\parallel, p_\perp) = \frac{1}{f} \frac{\partial f}{\partial p_\perp}
	\end{equation*}
which is equivalent to equation (\ref{eq:log:derivative}).

Note that, as $N \to \infty$, $\Delta v \to 0$, so $A,B \to 0$ and $\delta \to 1$, such that the effect of $\delta$-splitting is negligible. Of course, as $N \to \infty$, the differences in $f$ between two neighbouring nodes becomes negligible, and therefore $\delta$-splitting is no longer required. The introduction of this $\delta$-splitting technique therefore acts to improve the solution for coarse grids, while $\delta \approx 1$ for sufficiently fine grids.

\subsection{Numerical stability}
The $\delta$-splitting method has been introduced to ensure the steady-state solution of the electron distribution function under the effect of only the Fokker-Planck collision operator is a Maxwellian distribution, due to the distribution being a strongly (exponentially) varying function of $v$. The introduction of $\delta$, however, leads to a stability issue. Consider, for example, if $f$ is a Maxwellian,
	\begin{equation*}
	f \sim \exp{(-v^2)}
	\end{equation*}
and therefore
	\begin{equation*}
	g \sim -v
	\end{equation*}

In the limit $v \to \infty$, we have $A \to 0$, but $B \to \mp \infty$, such that $\delta \to \infty$. Numerically, this introduces problems, as large terms lead to instabilities by creating ill-conditioned matrices. The value of $\delta$ must therefore be limited to some maximum $\delta_\textrm{max}$. Fortunately there is the competing effect that $\delta \to 1$ for increasing $N$, such that, if $N$ is large enough, the value of $\delta_\textrm{max}$ is irrelevant as $\delta$ is always small enough for a stable solution, while the cutoff value $\delta_\textrm{max}$ only comes into effect at large $v$ where there are very few particles.

As an example, consider the case of an initial Maxwellian distribution with $T_e = 20 \, \text{eV}$ colliding with a fixed background distribution at $T_b = 10 \, \text{eV}$ with density $n_e = 10^{14} \, \text{m}^{-3}$. The number of grid points ($N = 150$) is chosen in order to demonstrate the effect of $\delta_\textrm{max}$, as increasing the number of grid points can eliminate the need for $\delta_\textrm{max}$.

\begin{figure}[!hbt]
	\centering
		\subfloat[]{%
			\includegraphics[width=0.33\textwidth]{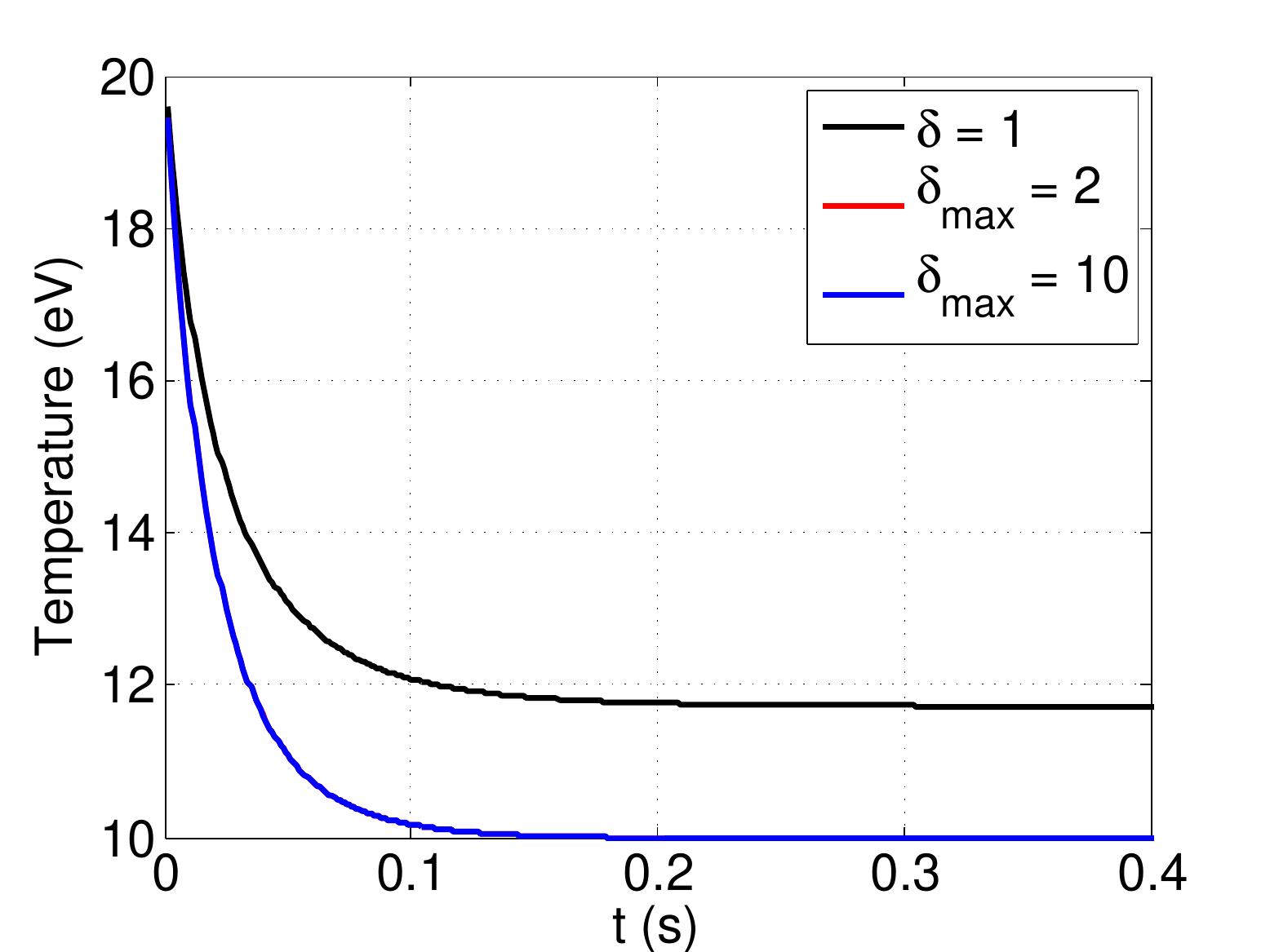}}
		\hfill
		\subfloat[]{%
			\includegraphics[width=0.33\textwidth]{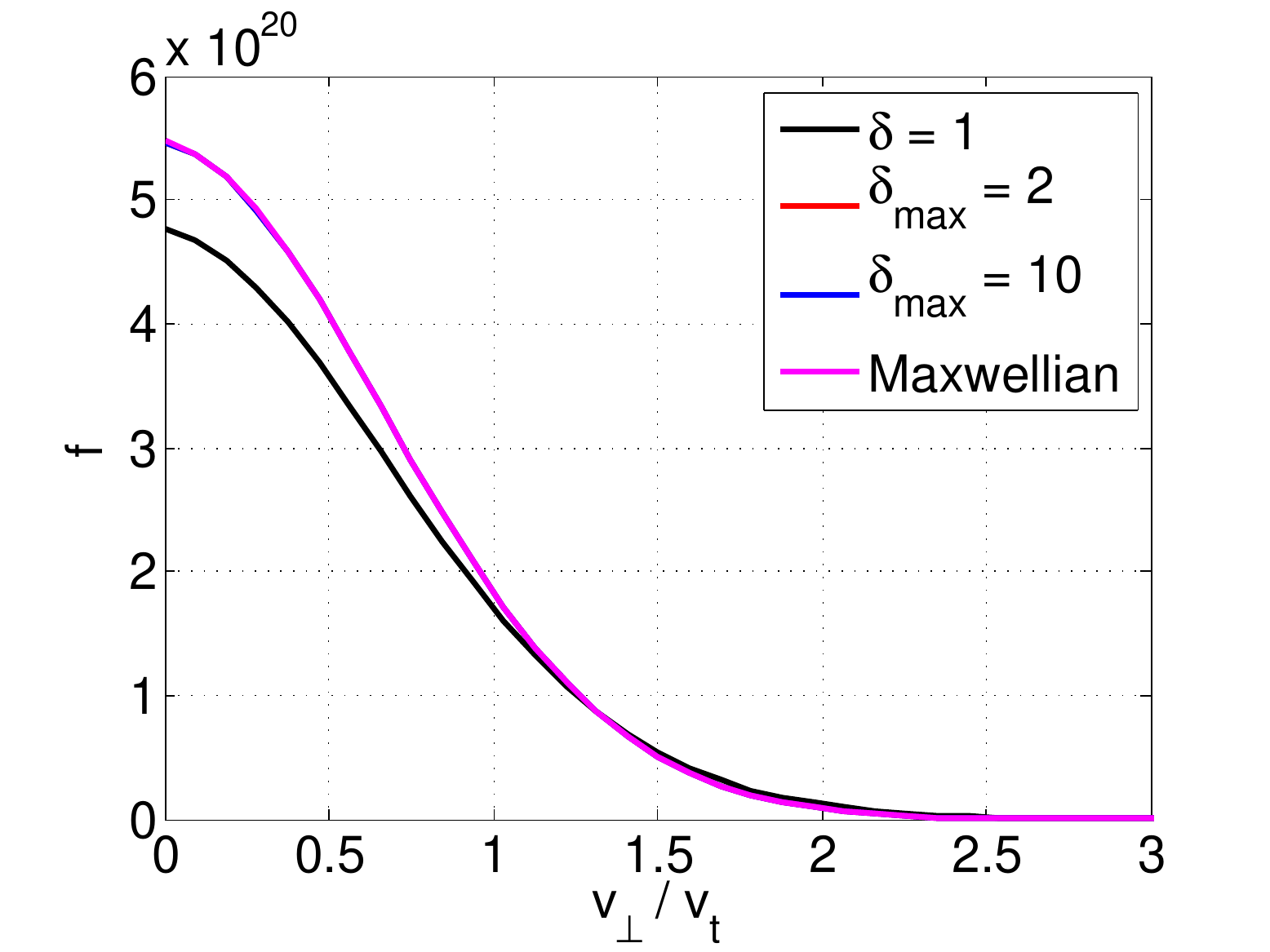}}
		\hfill
		\subfloat[]{%
			\includegraphics[width=0.33\textwidth]{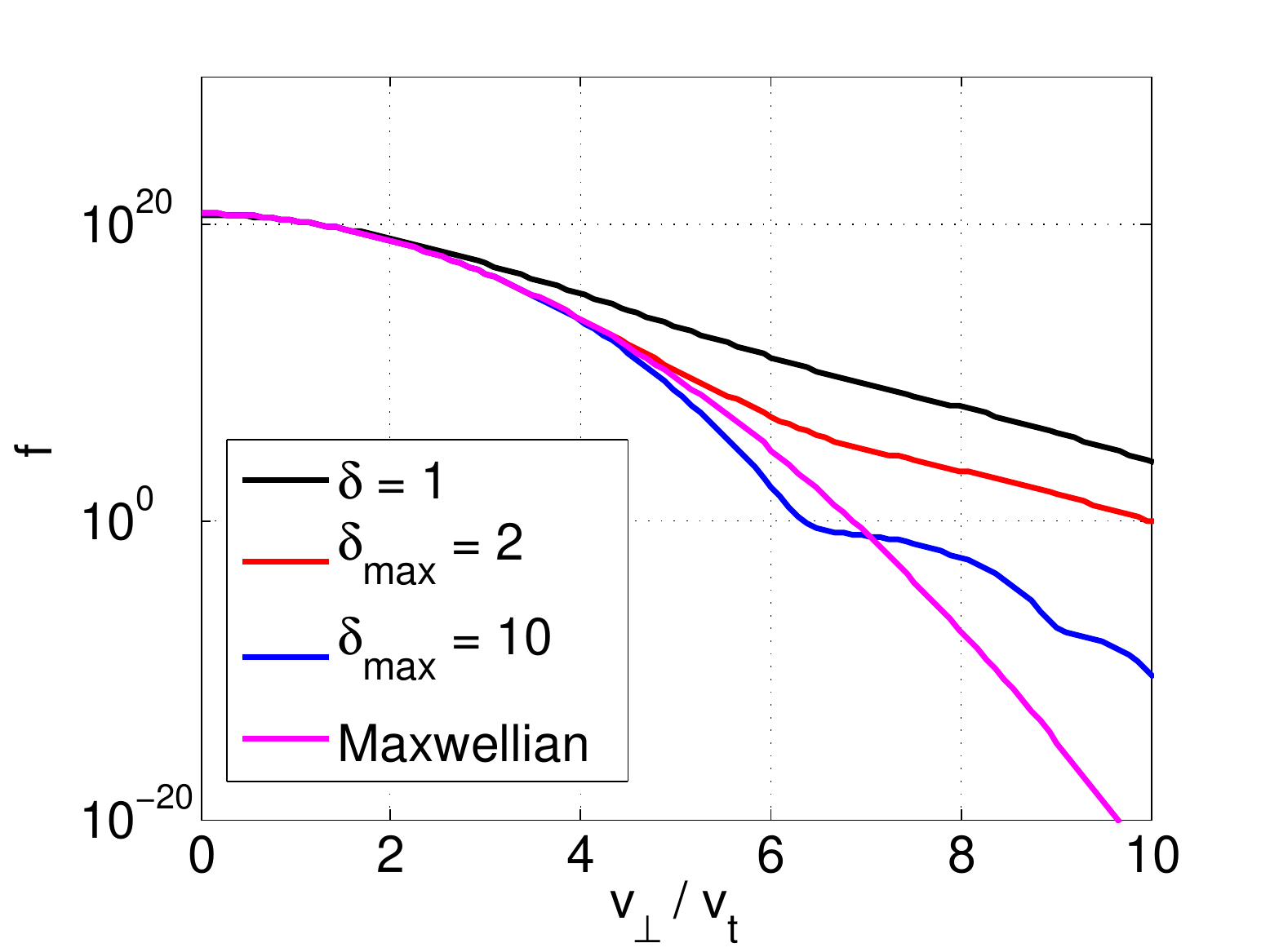}}
	\caption[]{(a) The time evolution of the temperature of an electron distribution function colliding with a fixed, background Maxwellian at $T_e = 10 \, \text{eV}$ and $n_e = 10^{14} \, \text{m}^{-3}$, for different choices of $\delta$ and $\delta_\textrm{max}$. There is no distinction between the temperature curves for $\delta_\text{max} = 2$ and $\delta_\text{max} = 10$. (b) The comparison of the steady-state distribution functions obtained shows no difference between $\delta_\text{max} = 2$, $\delta_\text{max} = 10$ and the background Maxwellian, except at large values of $v$, (c) shown on a log scale.}
	\label{fig:cylindrical:dmax}
	\end{figure}

Equilibrium is reached after $1 \, \text{s}$, and the distribution functions at this time, as well as the temperature evolution, are compared in figure \ref{fig:cylindrical:dmax}. Without $\delta$-splitting ($\delta = 1$), the wrong equilibrium temperature is reached, while for $\delta_\textrm{max} = 2$ and for $\delta_\textrm{max} = 10$ there are no differences in the temperature evolution, and the correct equilibrium temperature is reached, as expected, since $\delta$-splitting ensures that the solution is a Maxwellian.

Comparing the distribution functions to the background distribution shows the effect of $\delta_\textrm{max}$. Firstly, for $\delta = 1$ the final distribution is different from the background distribution, due to the wrong equilibrium temperature being reached. Comparing the $\delta_\textrm{max} = 2$ and $\delta_\textrm{max} = 10$ distributions to the background distribution, it is clear that there are only small differences for $v_\perp < 4 v_t$, where the majority of electrons are, and therefore the correct temperature is obtained. The differences are in the high $v$ tail, with $\delta_\textrm{max} = 2$ underestimating the relaxation, and $\delta_\textrm{max} = 10$ overestimating the relaxation.

Larger values of $\delta_\textrm{max}$ lead to ill-conditioned matrices as equilibrium is approached, so for numerical stability, the value of $\delta_\textrm{max}$ must be relatively small ($\delta \ge 1$ always). Of course, as $N \to \infty$ the value of $\delta \to 1$, and there will be no need for $\delta_\textrm{max}$. In practice, however, the grid will hardly ever be large enough to allow for this to happen, so it will be necessary to specify a value for $\delta_\textrm{max}$. Fortunately, this value will only impact regions of large $v$, where there are very few electrons which does not influence the low-order moments, such as temperature, and a relatively small value for $\delta_\textrm{max}$ suffices.

\subsection{Temperature equilibration}
Consider two electron distributions of equal density colliding with each other. The distributions will equilibrate according to
	\begin{equation}
	\frac{dT_a}{dt} =  \nu (T_b - T_a)
	\end{equation}
with $\nu$ the collision frequency, and
	\begin{equation*}
	\frac{dT_a}{dt} = -\frac{dT_b}{dt}
	\end{equation*}

The collision frequency is given by \cite{Callen_2006},
	\begin{equation}
	\label{eq:Callen:cylindrical}
	\nu = \frac{8}{3 \sqrt{\pi}} \left( \frac{e^2}{4\pi \varepsilon_0} \right)^2 \frac{4 \pi n_e \lambda}{m_e^2 \sqrt{\left( v_{t,a}^2 + v_{t,b}^2 \right)}}
	\end{equation}
where $v_{t,a}^2 = 2T_a/m_e$ is the thermal velocity of distribution $a$, and the Coulomb logarithm is taken to be constant $\lambda = 15$.

\begin{figure}[!hbt]
	\centering
		\hspace*{\fill}
		\subfloat[]{%
			\includegraphics[width=0.45\textwidth]{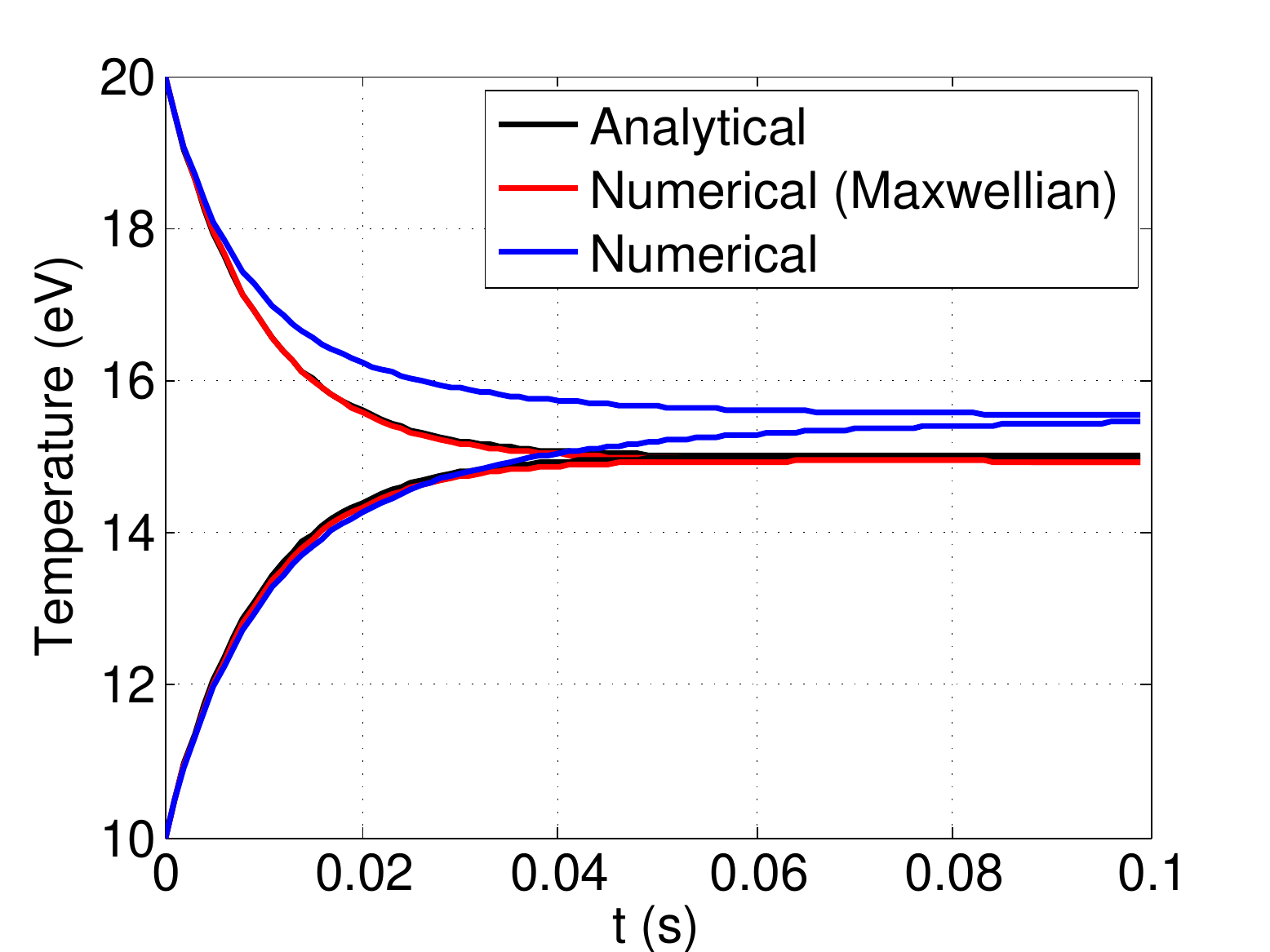}}
		\hfill
		\subfloat[]{%
			\includegraphics[width=0.45\textwidth]{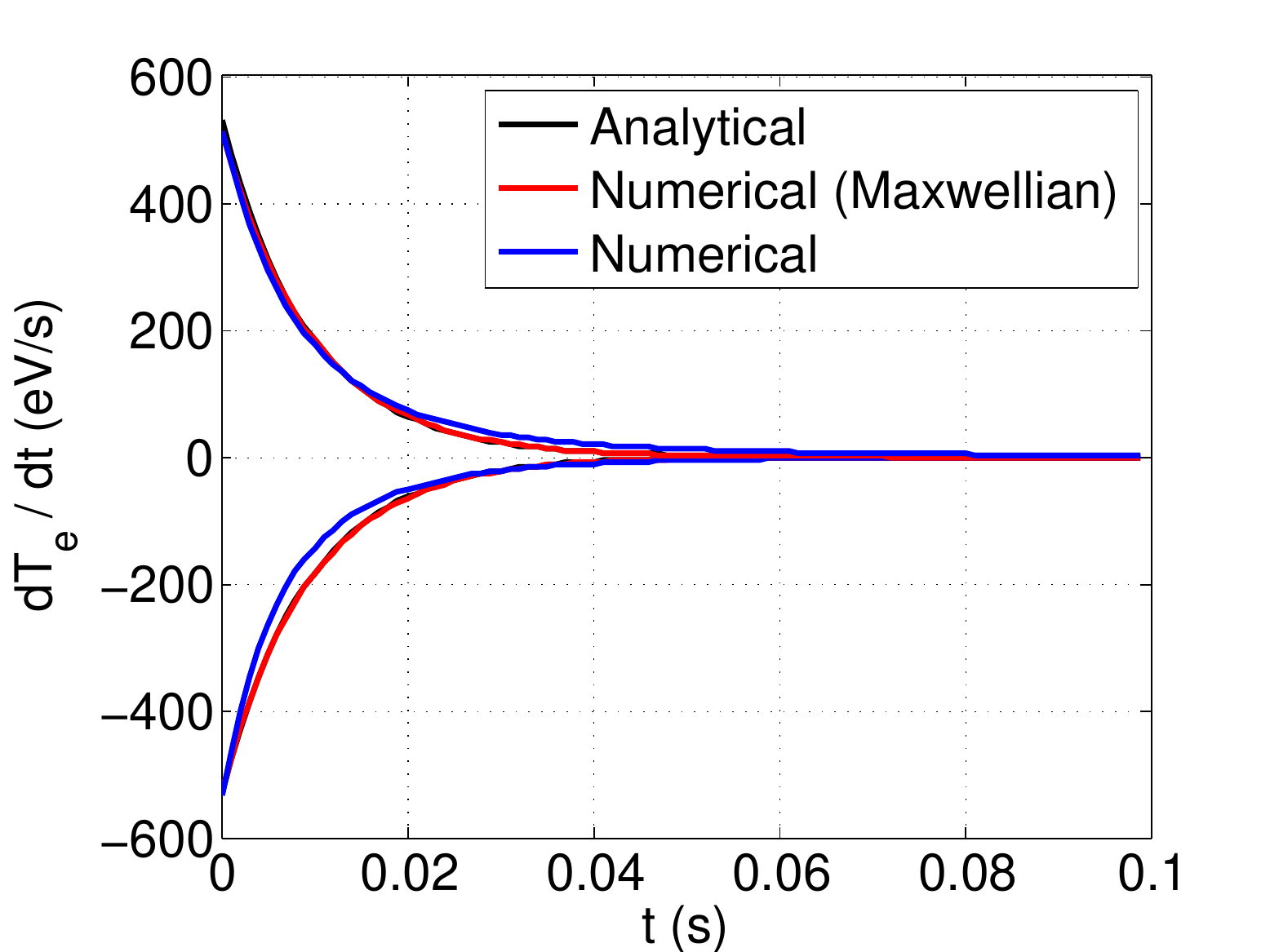}}
		\hspace*{\fill}
	\caption[]{The time evolution of (a) the temperature and (b) the temperature gradient for two distributions of density $n_e = 10^{14} \, \text{m}^{-3}$ colliding with each other, compared to the analytical formula. By constraining the distribution functions to be Maxwellian, agreement with the analytical formula is found, while the wrong equilibrium temperature is found otherwise.}
	\label{fig:equilibration:cylindrical}
	\end{figure}

Consider two distributions, with temperatures $T_a = 20 \, \textrm{eV}$ and $T_b = 10 \, \textrm{eV}$ and densities $n_a = n_b = 10^{14} \, \text{m}^{-3}$ colliding with each other. The resultant temperature evolution, when using $\delta$-splitting, is shown in figure \ref{fig:equilibration:cylindrical}. Of course, the assumption that both distributions collide with a background Maxwellian is not always true; colder electrons undergo more collisions than faster electrons, such that the distribution will not always be a Maxwellian. This results in the wrong equilibrium temperature, as the colder distribution heats up faster than the warmer distribution cools down, and energy is not conserved.

By constraining both distributions to always be Maxwellian (by replacing each distribution with a Maxwellian of the same temperature at each time step), the assumption of collisions with a background Maxwellian remains true, and the correct equilibrium temperature is reached at the predicted rate, while energy is conserved.

If the collision operators are calculated through the Rosenbluth potentials \cite{Rosenbluth_1957} or the Landau integrals \cite{Landau_1937}, it will no longer be necessary to constrain the distributions to be Maxwellian, as this constraint is only necessary to ensure the assumption of background Maxwellians is satisfied. The calculation of the collision operators from the distribution functions, however, adds additional numerical evaluations, and therefore is not considered here.

\subsection{Time evolution}
The treatment of the mixed derivatives introduces a non-linearity which can be solved through a Picard iteration. Alternatively, by Picard linearizing, an accurate solution can be obtained by using a smaller time step. To illustrate this, consider two initial Maxwellian distributions with $T_a = 20 \, \text{eV}$ and $T_b = 10 \, \text{eV}$ colliding with each other. Let $N = 150$ and $p_\textrm{max} = 45 \times 10^{-3} \, \textrm{MeV}/c$ such that $p_t = 10 \Delta p$, where $p_t = m_e v_t / c$ is the thermal momentum. The density is $n_e = 10^{14} \, \text{m}^{-3}$ such that the collision time $\tau \approx 20 \, \text{ms}$.

\begin{figure}[!hbt]
	\centering
		\subfloat[]{%
			\includegraphics[width=0.33\textwidth]{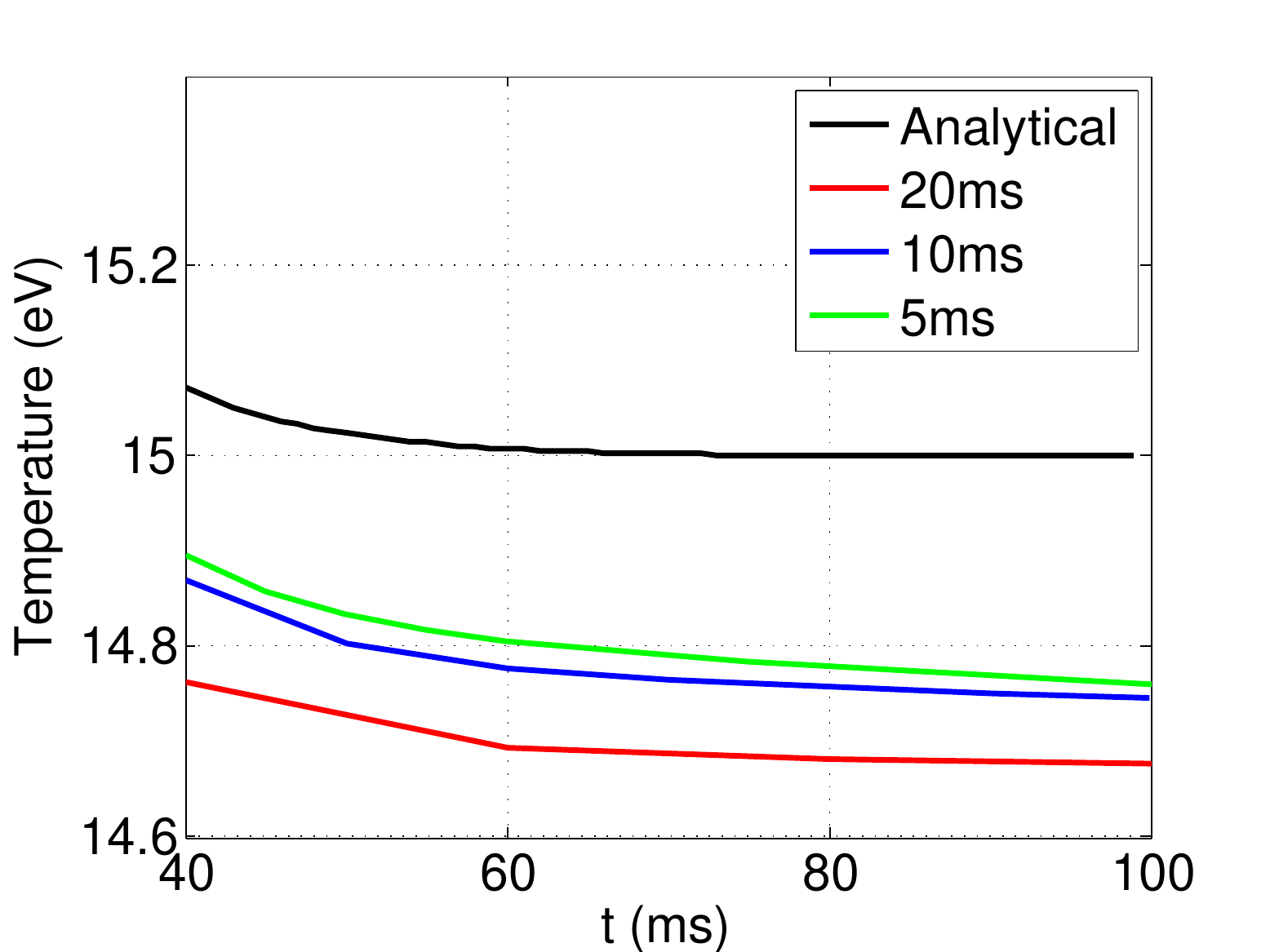}}
		\hfill
		\subfloat[]{%
			\includegraphics[width=0.33\textwidth]{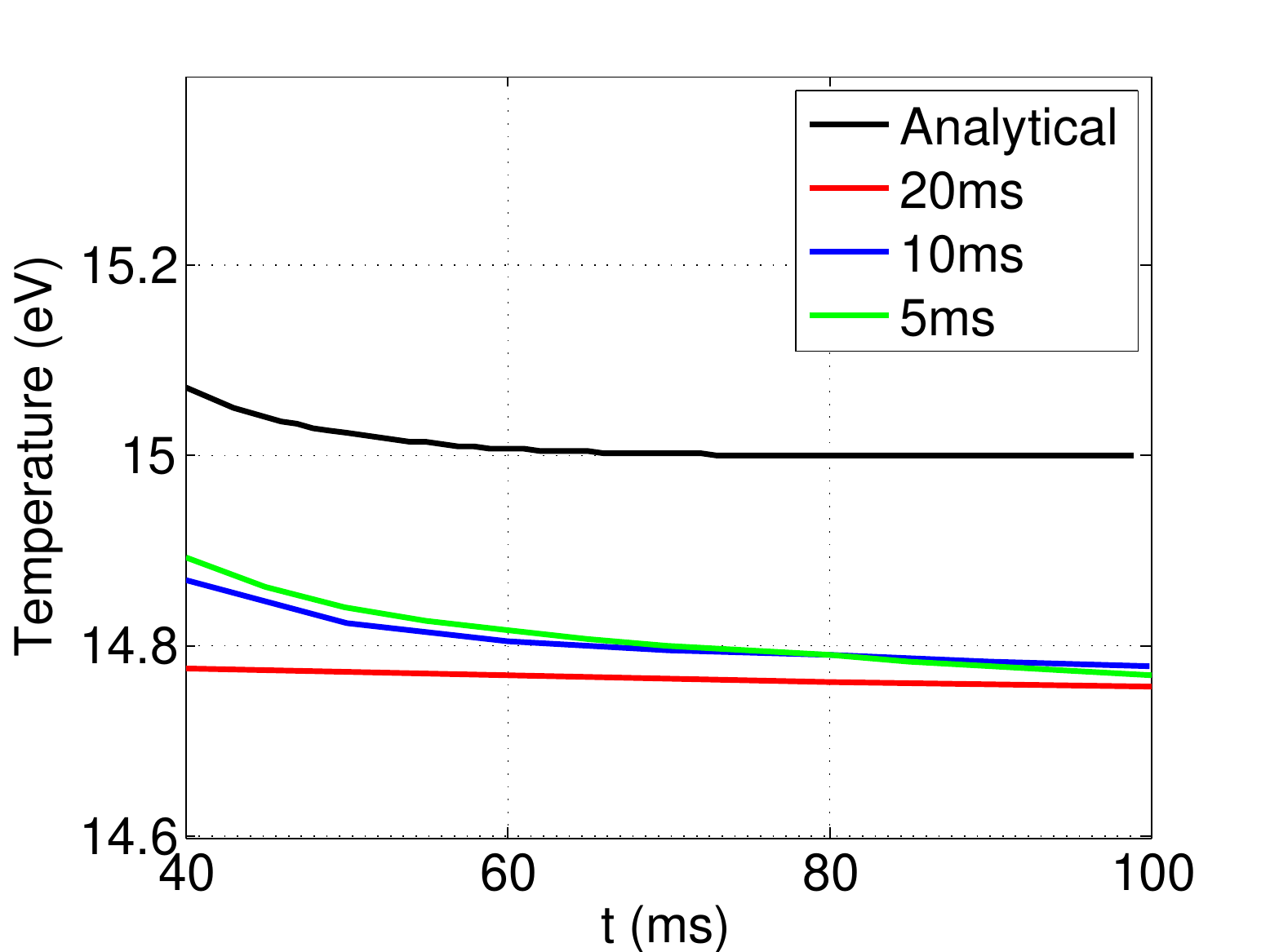}}
		\hfill
		\subfloat[]{%
			\includegraphics[width=0.33\textwidth]{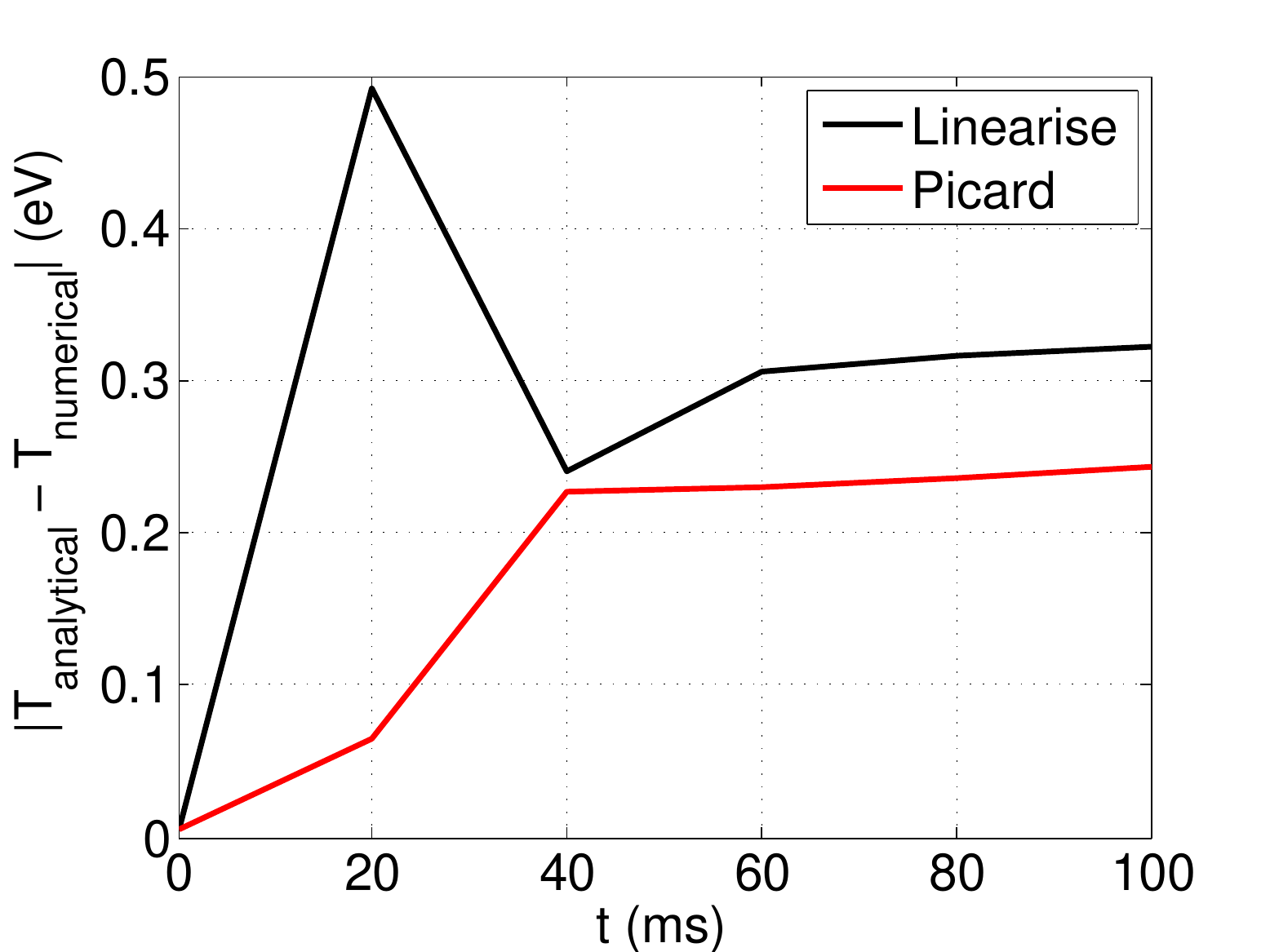}}
	\caption[]{The time evolution of the temperature of a distribution function at $T_a = 20 \, \text{eV}$ colliding with a distribution at $T_e = 10 \, \text{eV}$, both with density $n_e = 10^{14} \, \text{m}^{-3}$, for different choices of $\Delta t$ for (a) Picard linearizing, (b) Picard iterating, (c) and the comparison of the two methods for $\Delta t = 20 \, \text{ms}$ to the analytical temperature.}
	\label{fig:cylindrical:time}
	\end{figure}

The time evolution of the temperature for different values of $\Delta t$ is shown in figure \ref{fig:cylindrical:time}. As expected, convergence is achieved for larger time steps when solving the distribution through Picard iteration as compared to Picard linearizing. However, obtaining a solution through Picard iteration is computationally more expensive as we have to iterate over the solution $f^{n+1}$.

\section{Conclusion}
Lower-order numerical methods are less accurate, but generally more robust and reliable, than higher-order methods, while higher-order methods are more accurate, but also more complicated. In this paper we propose a scheme that focuses on improving the accuracy of lower-order methods, in particular with respect to the preservation of positivity, for solving two-dimensional advection-diffusion equations of the form
	\begin{equation*}
	\frac{\partial u}{\partial t} = \nabla \cdot \left( - \vec{a} u + \hat{k} \cdot \nabla u \right)
	\end{equation*}
where $u = u(x,y,t)$ is advected by the vector $\vec{a}(x,y,t)$ and diffused by the tensor $\hat{k}(x,y,t)$. Numerical solutions to these equations, in the absence of mixed derivatives, have been studied in detail \cite{Fazio_2009, Hundsdorfer_1995}, but solutions where mixed derivative terms are present have received much less attention. The scheme proposed in this paper allows the mixed derivative terms to be written as an advection-type equation, after which a lower-order positivity-preserving method can be applied. It was discussed by using an example and then applied to the Fokker-Planck collision operator in cylindrical coordinates. Compared to central finite-difference methods, the proposed scheme obtains the same order of accuracy, with the added advantage of the solution being non-negative.



In our text core, the Fokker-Planck collision operator is approximated under the assumption of local thermal equilibrium, which introduces an averaging parameter to ensure an accurate steady-state distribution. The solution is tested with the thermal equilibration of two colliding Maxwellian distributions and compares well with the theoretically predicted rate. The scheme conserves particle number and preserves positivity, but only conserves energy and agrees with the theoretical equilibration rate if the distributions are constrained to be Maxwellian, as the collision operators are calculated under the assumption of a background Maxwellian.

The treatment of the mixed derivatives introduces a non-linearity, which can be solved through Picard iteration. This iteration allows for larger time steps to be taken, but also increases the computational time. For the considered example of studying the thermal equilibration of two electron distribution functions under the Fokker-Planck collision operator, Picard linearizing provides an accurate approximation to the time evolution when the time-step is shorter than the collision time.


In conclusion, the scheme proposed in this paper is ideal when requiring a positive solution under the conservation of particle number or flux, when a small time step is used such that the non-linearity in the mixed derivatives can be solved by Picard linearizing. Although higher-order methods and flux limiting can be used to obtain more accurate solutions, the proposed scheme uses lower-order methods to ensure a reliable and robust method which is less complicated than higher-order methods. The $\delta$-splitting method can, in general, be applied to any advection-diffusion problem to ensure the correct steady-state solution. We show that, for equilibration under the Fokker-Planck collision operator, the scheme ensures the correct equilibrium distribution while conserving particle number and preserving positivity, while energy can be conserved by calculating the collision operators from the Rosenbluth potentials \cite{Rosenbluth_1957} or the Landau integral \cite{Landau_1937}.

\section*{Acknowledgements}

This work was funded by the RCUK Energy Programme under grant EP/P012450/1, and the University of York, through the Department of Physics and the WW Smith Fund. This work made use of the York Advanced Research Computing Cluster (YARCC) at the University of York.

\appendix


\section{Positivity-preserving solution to the advection equation}
\label{app:Hundsdorfer}
Consider the linear advection equation,
	\begin{equation}
	\label{eq:app:advection}
	u_t + (a u)_x = 0
	\end{equation}
which can be discretized,
	\begin{equation*}
	(au)_x = \frac{1}{\Delta x} ( F_{i+1/2} - F_{i-1/2} )
	\end{equation*}
such that, for $a > 0$,
	\begin{equation}
		\begin{aligned}
	F_{i+1/2} &= a_{i+1/2} \bigg( u_i + \frac{1}{2} \phi_{i+1/2} (u_i - u_{i-1}) \bigg) \\
	F_{i-1/2} &= a_{i-1/2} \bigg( u_{i-1} + \frac{1}{2} \phi_{i-1/2} (u_{i-1} - u_{i-2}) \bigg)
		\end{aligned}
	\end{equation}
where the function $\phi_{i\pm1/2}$ is the flux limiter. If $\phi = 0$, the scheme is first-order accurate, and known as the donor-cell upwind (DCU) scheme. For $\phi = 1$, the scheme is second-order accurate, but does not conserve extrema \cite{Fazio_2009}. First-order flux-conserving schemes, such as the DCU scheme, always preserves monotonicity, but higher-order schemes only preserves monotonicity with the help of flux limiters.

The flux limiter ensures that one gets the best of both second- and first-order methods. If the second-order scheme creates false extrema, then the first-order scheme is used. In order to ensure the best of both methods, the flux limiter is introduced. There exists a number of flux limiter functions (see, for example \cite{Fazio_2009}), but in this work we use the flux limiter introduced by Hundsdorfer \cite{Hundsdorfer_1995}, which gives
	\begin{equation}
	\phi_{i+1/2} = \text{max}(0, \text{min}(2r,\text{min}(2,K(r))))
	\end{equation}
where
	\begin{equation}
	r_{i+1/2} = \frac{u_{i+1} - u_i}{u_i - u_{i-1}}
	\end{equation}
and
	\begin{equation}
	K(r) = \frac{1 + 2r}{3}
	\end{equation}
which provides a second-order accurate solution to the advection equation (\ref{eq:app:advection}) that preserves positivity in two dimensions.

\end{document}